# V-V BOND-LENGTH FLUCTUATIONS IN $VO_x$

J. B. Goodenough, F. Rivadulla, E. Winkler, and J.-S. Zhou

*Texas Materials Institute, ME/MS & E, ETC 9.102, The University of Texas at Austin, 1 University Station, C2201, Austin, TX 78712*

## ABSTRACT

We report a significantly stronger suppression of the phonon contribution to the thermal conductivity in $VO_x$ than can be accounted for by disorder of the 16 % atomic vacancies present in VO. Since the transition from localized to itinerant electronic behavior is first-order and has been shown to be characterized by bond-length fluctuations in several transition-metal oxides with the perovskite structure, we propose that cooperative V-V bond-length fluctuations play a role in VO similar to the M-O bond-length fluctuations in the perovskites. This model is able to account for the strong suppression of the thermal conductivity, the existence of a pseudogap confirmed by thermoelectric power, an anomalously large Debye-Waller factor, the temperature dependence of the magnetic susceptibility, and the inability to order atomic vacancies in VO.

## INTRODUCTION

Metallic TiO and VO each have the rock-salt structure with about 16 % cation and anion vacancies[1]. In these oxides, π-bonding t orbitals of the octahedral site $3d^2$ and $3d^3$ manifolds provide strong metal-metal (M-M) interactions across shared octahedral-site edges. The vacancies appear to be introduced in order to decrease the M-M separation so as to broaden the narrow π* band of t-orbital parentage, thereby stabilizing the electrons in this band[2]. Metallic NbO has 25 % cation and anion vacancies that are ordered[3]. The vacancies also order in TiO,[4] but no vacancy ordering has ever been reported for VO. Why the vacancies do not order in VO given the large Coulomb interaction between them has remained an unresolved question. In this paper we call attention to several evidences that VO approaches the transition from localized to itinerant electronic behavior from the itinerant-electron side, and we report experiments designed to probe whether the lack of vacancy ordering could be due to bond-length fluctuations induced by cooperative vanadium-atom displacements separating regions of longer and shorter V-V bonding.

We were motivated by experiments on transition-metal perovskite oxides at the crossover from localized to itinerant electronic behavior[5]. Inoue *et al.*[6] have provided direct evidence, from photoemission spectra (PES) taken on the metallic system $Sr_{1-x}Ca_xVO_3$, for the coexistence of itinerant (coherent) electrons in a partially filled conduction band and strongly correlated (incoherent) electrons in a lower Hubbard band. These spectra were interpreted to be direct evidence of strong-correlation fluctuations in an itinerant-electron matrix, and the PES data allowed identification of an indirect signature of these fluctuations in the pressure dependence of the thermoelectric power $\alpha(T)$[5]. The transition from Pauli paramagnetism toward Curie-Weiss paramagnetism in $La_{1-x}Nd_xCuO_3$[7] or $La_{1-x}Sr_xTiO_3$[8] could then be understood as the growth of a phase with strongly correlated electrons at the expense of a Pauli paramagnetic itinerant-electron phase. These two-phase fluctuations are not detectable by a conventional diffraction experiment, which only show an enhanced Debye-Waller parameter.

A logical next step is to ask whether two-phase fluctuations or a global first-order phase-change occurs at the crossover from localized to itinerant electronic behavior in systems where M-M interactions dominate or compete with M-O-M interactions. The most promising place to look for two-phase fluctuations is where the structure tends to frustrate long-range order into a charge density wave (CDW) as in the vanadium spinels and VO.

## EXPERIMENTAL

Banus, Reed, and Strauss[9] have reported the electrical resistivity and magnetic properties of polycrystalline $VO_x$ and $TiO_x$. Because of the roughly 16 % cation and anion vacancies at x = 1, both oxides can be prepared over a large compositional range ($0.70 < x < 1.32$). Polycrystalline $VO_x$ and $TiO_x$ have been generally synthesized by arc melting and casting. Single crystals could only be grown near the highest stable oxidation state corresponding to $1.27 \leq x \leq 1.32$. Arc melting and casting presents experimental difficulties that have resulted in disagreement between data published before the work of Banus, Reed and Strauss[9].

We used a different synthetic route to prepare $VO_x$; it is much more accessible than arc melting, and it provides high-density, single-phase polycrystalline materials over the whole homogeneity range. High purity vanadium metal and $V_2O_3$ were mixed in stoichiometric proportions, according to the desired value of x. The powders were ground, mixed, and pressed into pellets in an Ar atmosphere before being transferred and sealed into a silica tube that had been evacuated down to $P \approx 10^{-5}$-$10^{-4}$ Torr. The ampoules were annealed at 1100ºC for 24 h and quenched into an ice-water bath. The stability of VO at low temperatures is doubtful; disproportionation reactions have been reported[10], which is already an indication that two-phase fluctuations may appear below room temperature in quenched samples. Quenching from high temperature avoids the disproportionation problem, which was probably the origin of the metal-insulator transition attributed to VO in the old literature[11].

The quenched pellets were polished to remove any trace of $V_3Si$ from reaction with the tube[12], ground and cold-pressed at $16\times10^3$ kg/cm$^2$ before being again sealed in evacuated silica tubes and refired at 1000ºC for 24 h. After this treatment the pellets were polished and the x-ray patterns showed only very narrow peaks of the single-phase cubic (Fm-3m) material (Fig. 1). The lattice parameters for different x are in perfect agreement with the literature values.

The oxygen/metal ratio x in $VO_x$ was determined by thermogravimetric analysis (TGA). Powdered samples were calcined in oxygen at 1ºC/min up to 650ºC and held for 16 h. After this time, oxidation to $V_2O_5$ was complete; no further weight gain was detected, indicating complete combustion of the monoxide.

The grain sizes of the samples determined by optical microscopy typically ranged between 5 and 20 microns.

Monoclinic $TiO_x$ (space group *A2/m*) was provided by Alfa (99.5 %). TGA analysis revealed x=0.99(1). The powder was pressed, sealed in evacuated silica tubes and annealed at 1100ºC for 24 h. After this treatment, quenching in ice-water resulted in cubic (Fm-3m) TiO while slow cooling (holding the sample at 900ºC for at least 48 h.) gave the monoclinic phase (Fig. 1). Both TiO samples were stoichiometrically identical, but the monoclinic phase has an ordered array of vacant lattice sites: half of the Ti and half of the O atoms are missing alternately in every third (110) plane[4].

All attempts to order the vacancies in the case of $VO_x$ were unsuccessful.

Magnetic measurements with small field strengths (between 10 and 50 Oe) down to 5 K were done in a SQUID magnetometer (Quantum Design). Thermal conductivity was measured with a steady-state method; the temperature gradient was controlled to be less than 1% of the base temperature.

## RESULTS AND DISCUSSION

Fig. 2 shows the temperature dependence of the thermoelectric power, $\alpha(T)$, for both TiO and $VO_x$. It is small and varied linearly above ≈ 120 K for all samples, as in a meta[13]. The phonon-drag effect on the $\alpha(T)$ has not been observed in the temperature range of our experiments.

In TiO ($3d^2$), the $\pi^*$ band of width $W_\pi$ is one-third filled and the on-site electrostatic energy is $U_\pi < W_\pi$, so $\alpha(T)$ varies little with x in $TiO_x$ and remains negative for all x. On the other hand, $\alpha(T)$ for $VO_x$ ($3d^3$) changes from negative to positive at x = 1.02 as x increases. This behavior, which was already noted in an earlier study[1], signals the opening of a pseudogap at the energy of half-filling of the $\pi^*$ band. The Fermi energy $\varepsilon_F$ lies above half-filling of the $\pi^*$ band for x < 1, which introduces electrons into the upper Hubbard band; for x > 1, $\varepsilon_F$ falls below half-filling, which introduces holes into the lower Hubbard band. In the localized-electron limit for an octahedral-site $t^3e^0$ configuration, the effective $U_\pi$ is augmented by Hund's intraatomic exchange splitting $\Delta_{ex}$ of states of different spin and $U_{\pi\sigma}$ is augmented by the crystal-field splitting $\Delta_c$. Therefore, a $W_\pi \approx U$ is approached in VO. Earlier recognition[2] that this was the origin of the pseudogap in VO could provide no clear model of how the pseudogap develops on the approach to the Mott-Hubbard transition from the itinerant-electron side since it is assumed in the Mott-Hubbard model that the crossover to localized-electron behavior occurs within a single electronic phase[3]. The fluctuating two-phase model provides a clear physical basis for the opening of the pseudogap as discussed above.

A careful inspection of the x-ray spectra of our samples gave no indication of extra peaks or anomalous peak broadening that would signal the presence of static phase segregation. Therefore, we conclude that the transfer of spectral weight to Hubbard bands split by an energy gap is caused by dynamic phase fluctuations. A careful x-ray study by Morinaga and Cohen[14] of several $VO_x$ compositions supports this conclusion. They found an anomalously large Debye-Waller factor of 1.1 Å$^2$ for VO compared to 0.56-0.8 Å$^2$ for $TiO_x$ and 0.79 Å$^2$ for $Fe_{1.9}O$ where disordered vacancies are also present. The larger Debye-Waller factor in $VO_x$ may be associated with bond-length fluctuations in addition to static atomic displacements of atoms neighboring atomic vacancies.

The magnetic susceptibility of monoclinic TiO was found to be small and essentially temperature-independent below about 280 K, see inset of figure 3. The change in slope of $\chi^{-1}(T)$ near 280 K is even more pronounced in cubic TiO where the temperature dependence of $\chi(T)$ below 280 K is also somewhat greater. An anomaly is also observed in the thermal conductivity around this temperature, as we will show later. A probable cause of this effect is trapping of two electrons at an oxygen vacancy and the thermal excitation of one of them to the itinerant-electron band to leave a localized spin behind. In contrast, the $\chi^{-1}(T)$ curves for $VO_x$ had a strong temperature dependence as is illustrated in Fig. 3 for $VO_{1.065}$. The solid line is a fit to the expression

$$\chi = \chi_0 + \frac{C}{T+\theta} \qquad (1)$$

which corresponds to a Pauli paramagnetic component $\chi_o$ from an itinerant-electron phase and a Curie-Weiss component from strong-correlation fluctuations. From the Curie constant C, an effective magnetic moment $\mu_{eff}$ per vanadium atom was found to be 0.4 to 0.65 $\mu_B$ with a Weiss constant $-10\text{ K} \le \theta \le -3\text{ K}$ over the compositional range $1 \le x \le 1.16$. If we assume S = 3/2 at the vanadium with strongly correlated electrons, the measured $\mu_{eff}$ would correspond to 10 to 16 % of the vanadium in strong-correlation clusters, which is too small a fraction to percolate and stabilize a static antiferromagnetic order with a Weiss constant that reflects long-range antiferromagnetic interactions.

In order to test further the hypothesis of cooperative bond-length fluctuations in $VO_x$, we measured the temperature dependence of the lattice component of the thermal conductivity $\kappa_{ph}(T)$ for several compositions and compared the results with similar measurements on ordered and disordered TiO (Fig. 4).

Suppression of $\kappa_{ph}(T)$ proves to be an excellent indicator of the presence of bond-length fluctuations whether these fluctuations derive from disordered Jahn-Teller deformations[15], spin-lattice coupling in charge-transfer-gap paramagnets[16], or two-phase fluctuations[17].

The electronic component $\kappa_e(T)$ of the measured $\kappa(T) = \kappa_{ph}(T) + \kappa_e(T)$ was subtracted out by calculation from the Wiedermann-Franz law and the measured electronic conductivity. For a crystalline material[18]

$$\kappa_{ph} = \frac{1}{3} C_v v \lambda \qquad (2)$$

where $C_v$ is the specific heat, $v$ is the velocity of sound and $\lambda$ is the phonon mean free path, which is much smaller than the grain size of our samples. In an amorphous material, $\lambda$ is small and constant; in this case $\kappa(T)$ increases linearly with temperature due to an increasing carrier density, but its magnitude is strongly suppressed in comparison with that of crystalline materials that have phonons[19]. In a crystalline material, $\lambda$ decreases with increasing temperature, and $\kappa_{ph}$ normally exhibits a maximum at lower temperatures. In the presence of disordered atomic vacancies, $\lambda$ would be restricted to the distance between vacancies at lower temperatures, and the resulting suppression of $\kappa_{ph}$ should be similar in VO and cubic TiO where a comparable number of atomic vacancies are disordered in each. As can be seen in Fig. 4, all the $VO_x$ samples showed a larger suppression of $\kappa_{ph}(T)$ than that found in cubic TiO.

This added suppression cannot be due to the grain size; it also cannot be attributed to poor connectivity across grain boundaries since the resistivites of our samples were low (about $2\times10^{-3}$ Ωcm for $VO_{1.01}$, at 300 K) in agreement with those measured on arc-melted samples. The fact that $\kappa_{ph}(T)$ does not vary with x in spite of strong variations with x in vanadium interstitials[14], resistivity, thermoelectric power, and magnetic susceptibility rules out most other possibilities.

We conclude that the additional suppression of $\kappa_{ph}(T)$ in $VO_x$ signals the presence of bond-length fluctuations that limit $\lambda$ to a small, constant value as in an amorphous material.

As a test of this conclusion, we compare in Fig. 5 the $\kappa_{ph}(T)$ for the insulators MnO[20] and amorphous $SiO_2$[21] with $\kappa_{ph}(T)$ for $VO_{1.01}$ in a log-log plot. Whereas the $\kappa_{ph}(T)$ curve for MnO is typical of a crystalline solid, both VO and amorphous $SiO_2$ have a $\kappa_{ph}(T)$ that increases monotonically with temperature, a gradual rise following an intermediate plateau eventually saturating at the value of the minimum thermal conductivity for a glass; as developed by Cahill and Pohl[21], the minimum thermal conductivity for VO was obtained with $v = 4.5\times10^{-5}$ cm/s calculated from the Debye equation from $\theta_D = 614$ K[22].

We also call attention to the crystalline ferroelectrics like $Pb(Mg_{1/3}Nb_{2/3})O_3$ or $(Sr_{0.61}Ba_{0.39})Nb_2O_6$ that exhibit a diffuse dielectric transition at the ferroelectric transition temperature $T_C$ rather than a sharp transition as in $BaTiO_3$ and $Pb_5Ge_3O_{11}$[23]. Those with a diffuse transition have a glass-like thermal conductivity that has been attributed to the coexistence of two phases segregated on a small length scale, one becoming ferroelectric while the other remains paraelectric. These "glassy ferroelectrics" appear to contain bond-length fluctuations associated with two-phase fluctuations.

## CONCLUSIONS

We have compared $\kappa_{ph}(T)$ of VO with that of vacancy-disordered and vacancy ordered TiO as well as those of MnO and amorphous $SiO_2$. Whereas vacancy disorder suppresses $\kappa_{ph}(T)$ in TiO, there is an additional suppression of $\kappa_{ph}(T)$ in all the $VO_x$ compositions that makes their $\kappa_{ph}(T)$ approach the value and the temperature dependence found in amorphous materials. Since metallic VO contains 16 % atomic vacancies that, surprisingly, do not order and since VO approaches the Mott-Hubbard transition from the localized-electron side, it is a likely candidate for cooperative V-V bond-length fluctuations that segregate strong-correlation clusters within an itinerant-electron matrix. Moreover, bond-length fluctuations suppress the phonon component of the thermal conductivity. We therefore conclude that two-phase fluctuations are present in VO as a result of cooperative fluctuations of the V-V bond lengths. Two-phase fluctuations rather than a sharp first-order phase change may occur where there are strong perturbations of the periodic potential or where there is a frustration of long-range magnetic order as in the octahedral sites of a vanadium spinel. The existence of phase fluctuations could also account for the observation of a pseudogap in VO, an anomalously large Debye-Waller factor, a temperature-dependent magnetic susceptibility, and the absence of ordering of the atomic vacancies.

Identification of two different V-V bond lengths by pair-density-function analysis of pulsed-neutron-diffraction data, by photoemission spectroscopy, or by other fast experimental probes would provide direct proof of our deduction.

We would like to acknowledge support from the NSF, the Robert A. Welch Foundation, Houston Tx, and the Texas Center for Superconductivity at the University of Houston (TCSUH). F. R. would also like to thank the Fulbright foundation and the Ministery of Science of Spain for a postdoctoral fellowship.

## FIGURES

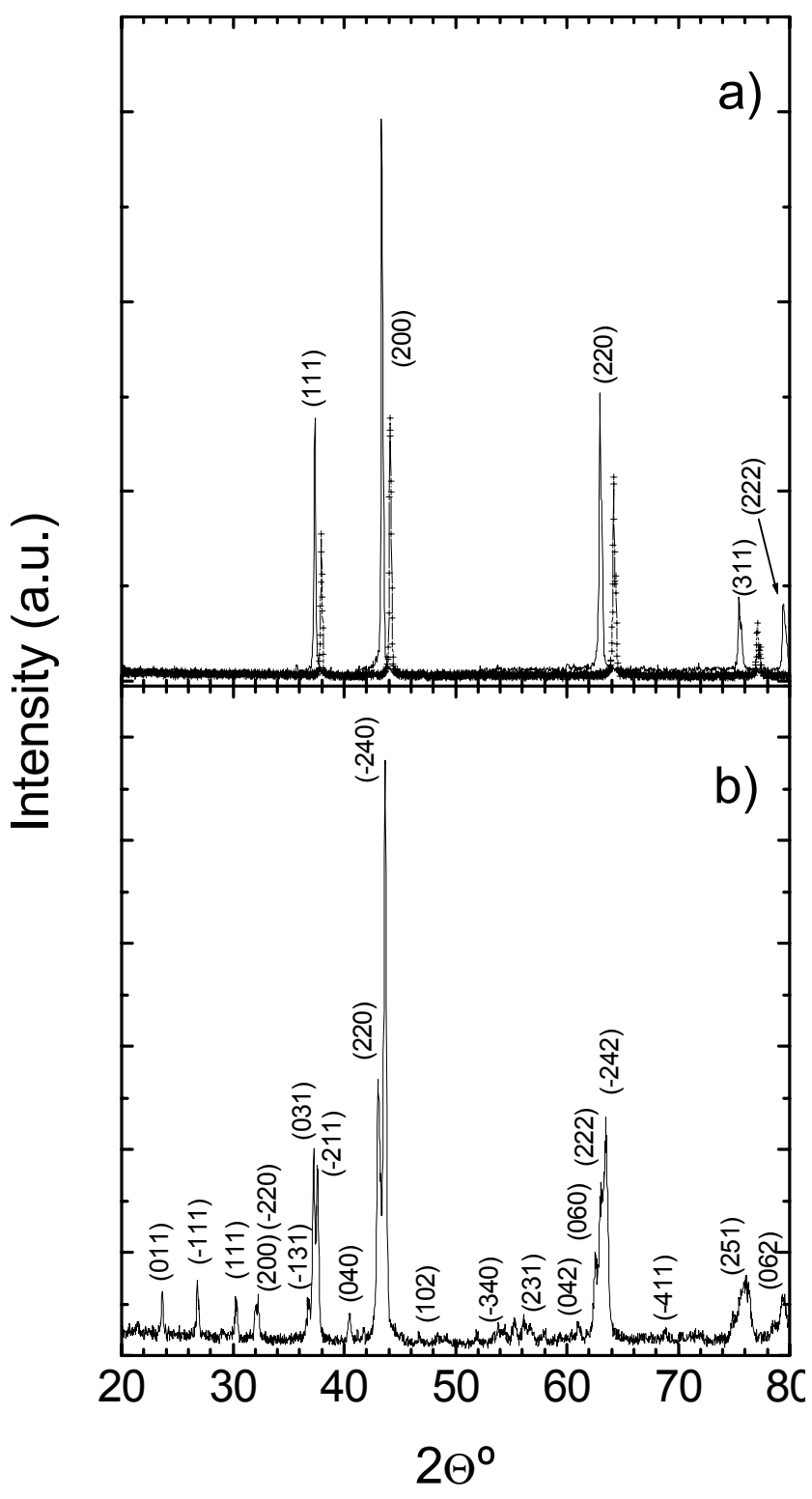

**FIG. 1.** (a) X-Ray powder patterns of cubic VO (crosses) and TiO (line). (b) Pattern for monoclinic TiO (a = 5.850(2), b = 9.342(2), c = 4.142(1), $\gamma$ = 107.53(3)º). Although all the peaks of (b) correspond to the monoclinic phase, only the main peaks are indexed for clarity.

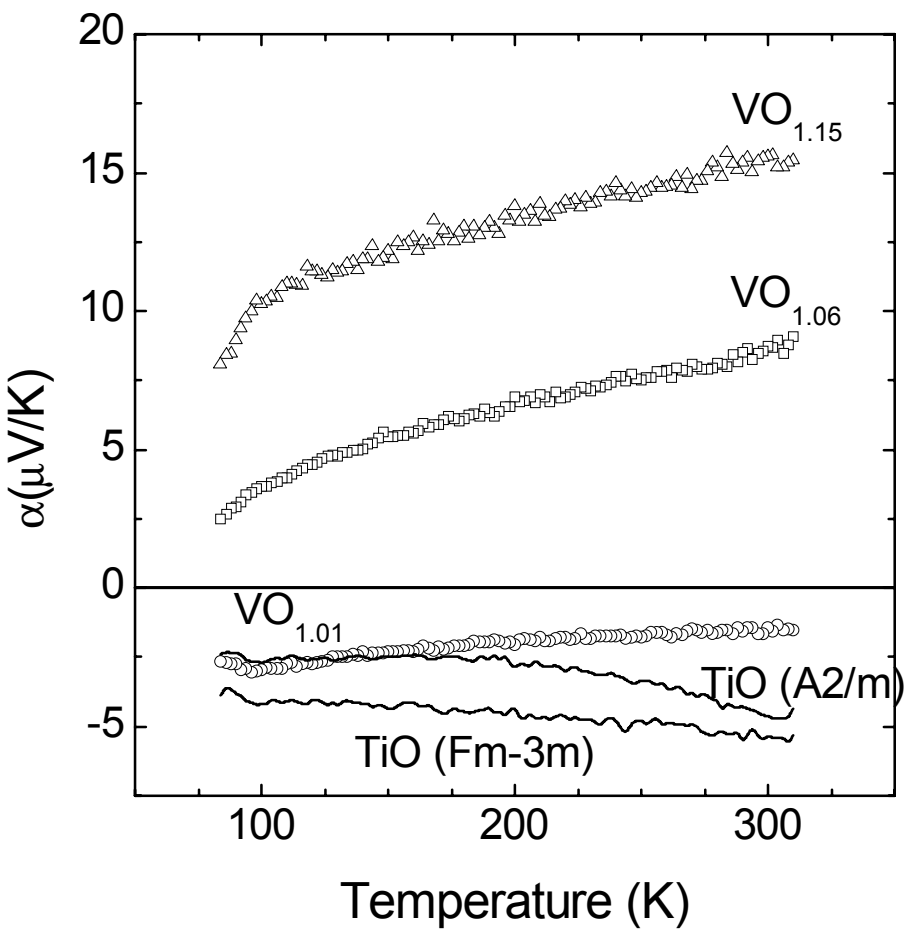

**FIG. 2.** Temperature dependence of the thermoelectric power for several compositions of $VO_x$ (symbols) and TiO (lines) with ordered and disordered vacancies.

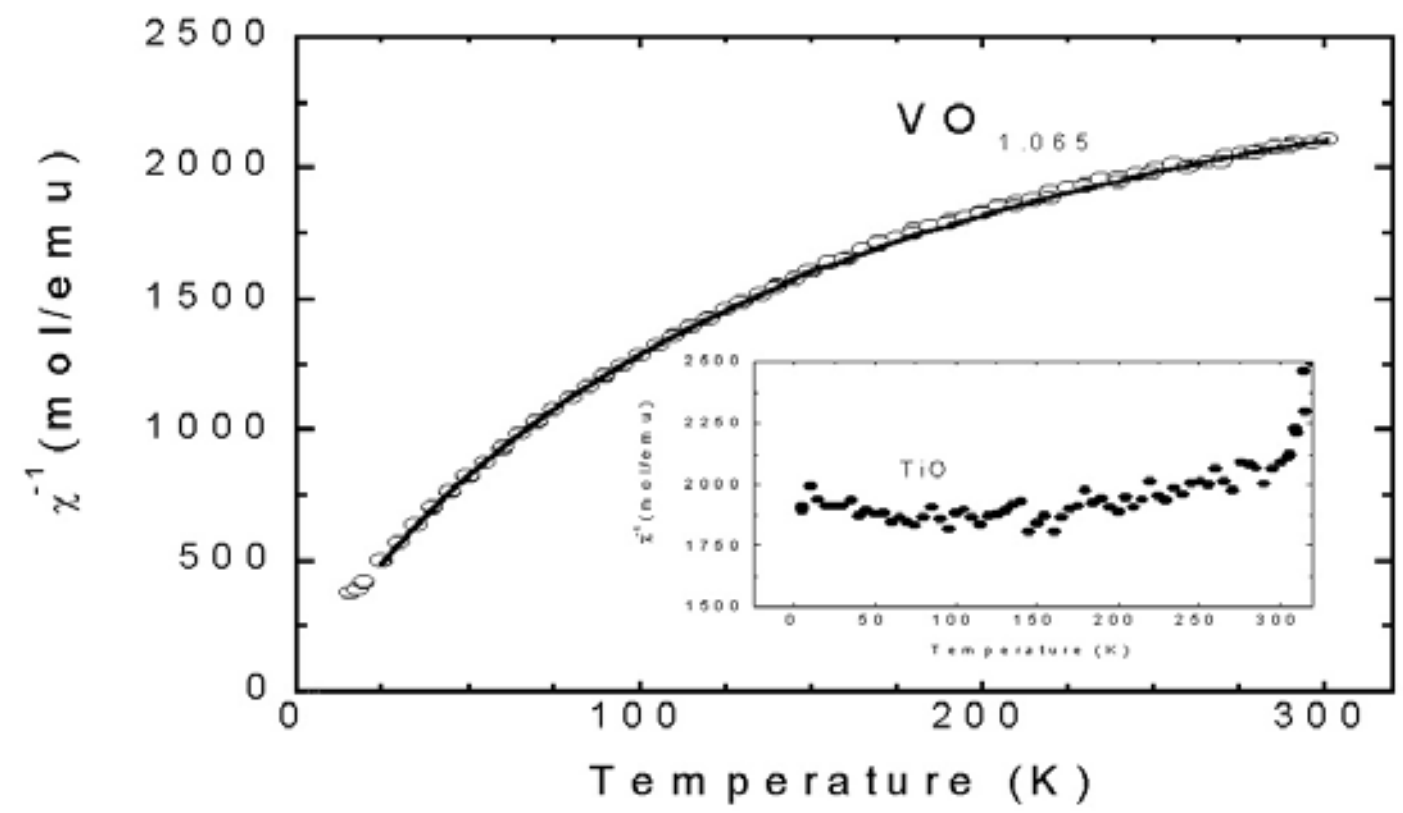

**FIG. 3.** Inverse molar susceptibility for $VO_{1.065}$. Inset: Inverse molar susceptibility for monoclinic TiO. The kink in the susceptibility is even more marked in the cubic phase, which also shows a more marked temperature dependence at lower temperatures. The kink is absent in the original, non-annealed monoclinic samples, which points to some disorder of the vacancies in our annealed monoclinic samples.

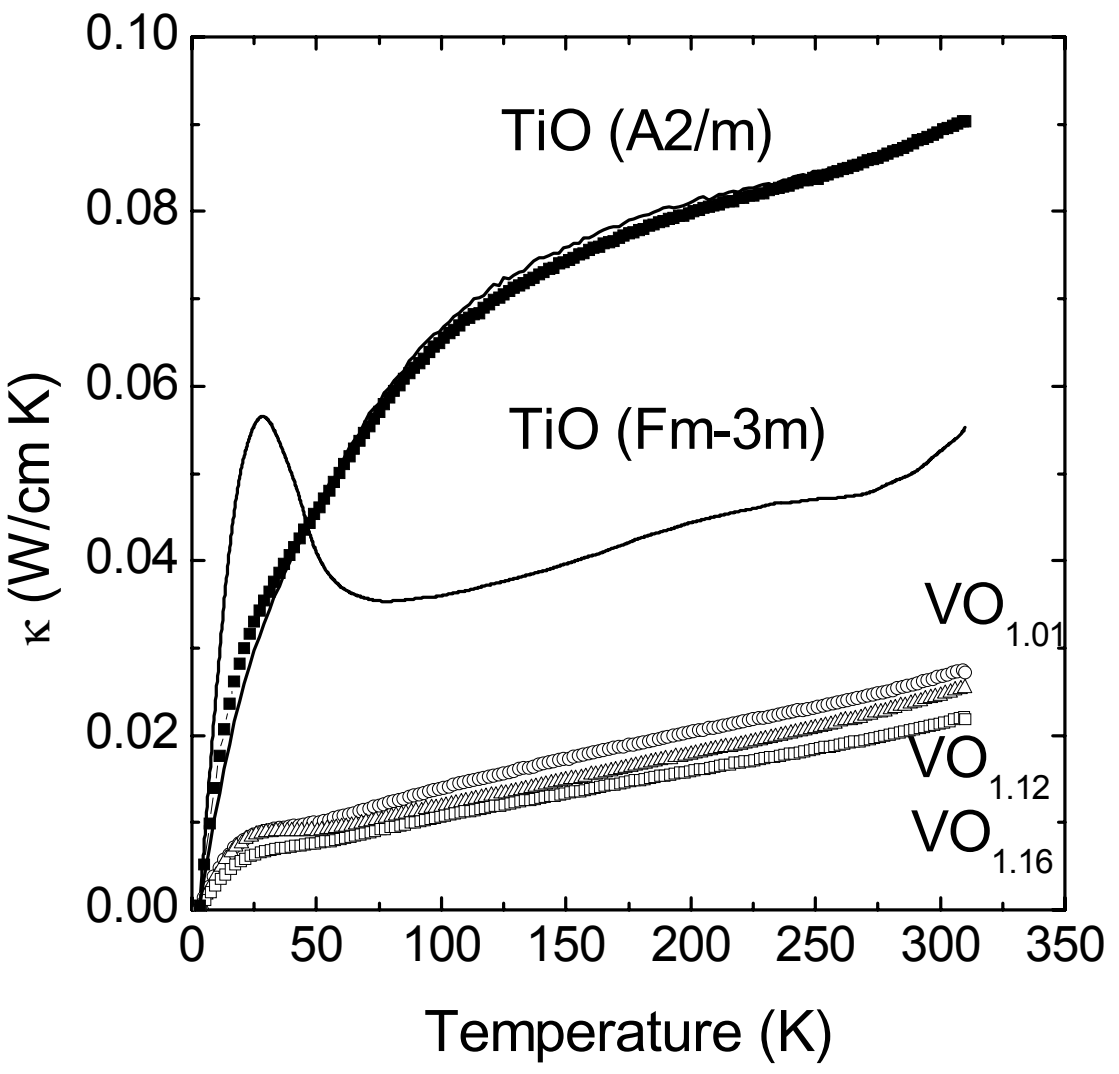

**FIG. 4.** Temperature dependence of the thermal conductivity, κ(T), for several compositions of $VO_x$ and for TiO with and without ordering of the vacancies.

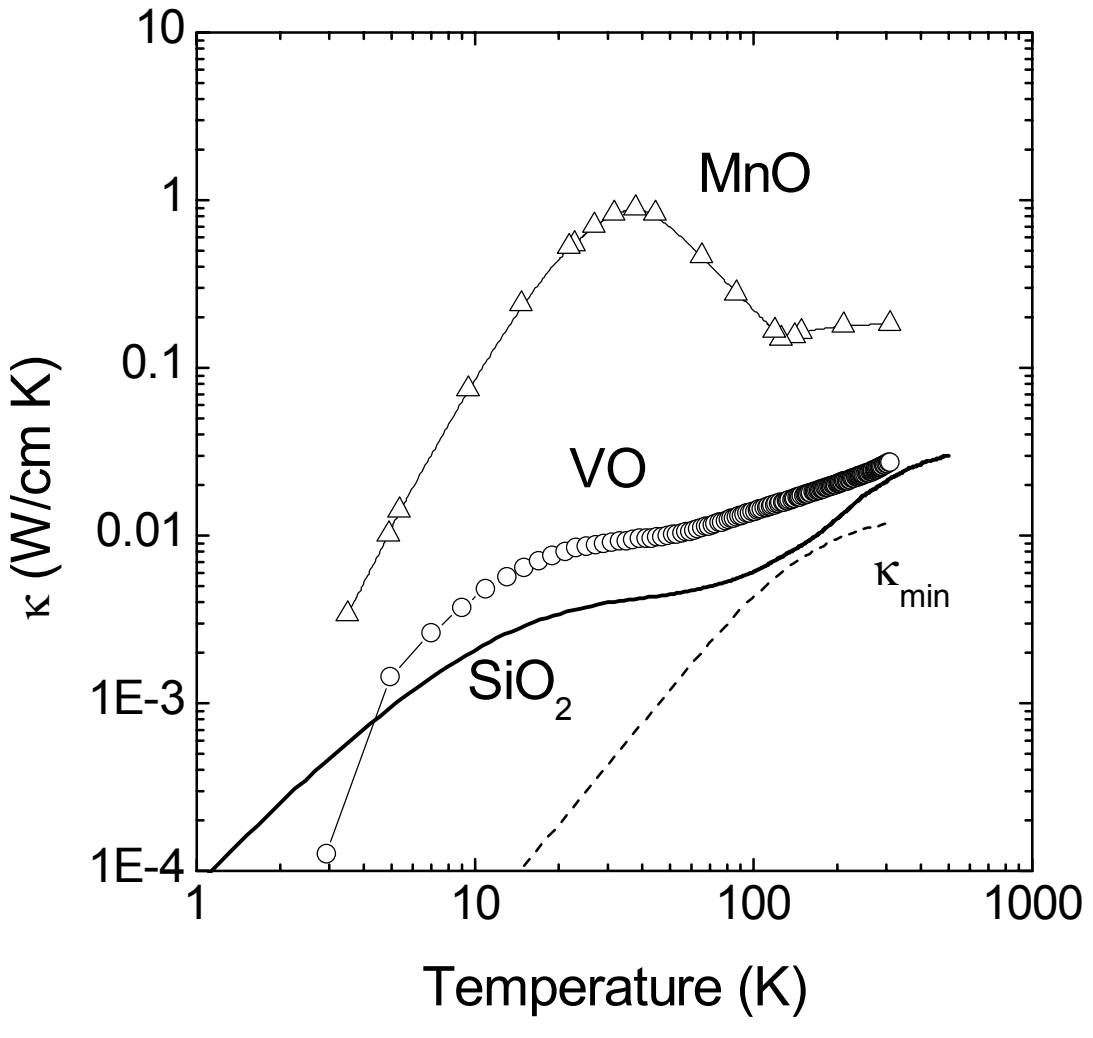

**FIG. 5.** Temperature dependence of the thermal conductivity of $VO_{1.01}$ (circles), MnO (triangles, from reference 20) and amorphous $SiO_2$ (solid line, from reference 21). The dashed line represents the calculated minimum thermal conductivity for VO.

## REFERENCES

[1] M. D. Banus, T. B. Reed, in *The Chemistry of Extended Defects in Non-Metallic Solids*, Ed. By L. Eyring and M. O'Keeffe, North-Holland, Amsterdam 1970, p. 488.
[2] J. B. Goodenough, Phys. Rev. B, **5**, 2764 (1972).
[3] J. B. Goodenough, Prog. Solid State Chem., **5** 145 (1971).
[4] D. Watanabe, J. R. Castles, A. Jostsons, and A. S. Malin, Acta Cryst. **23**, 307 (1967).
[5] J.-S. Zhou, J. B. Goodenough, Phys. Rev. B, **54**, 13393 (1996); T. Yoshida, A. Ino, T. Mizokawa, A. Fujimori, Y. Taguchi, T. Katsufuji, and Y. Tokura, Europhys. Lett. **59**, 258 (2002).
[6] I. H. Inoue I. Hase, Y. Aiura, A. Fujimori, Y. Haruyama, T. Maruyama, and Y. Nishihara, Phys. Rev. Lett., **74**, 2539 (1995)
[7] J.-S. Zhou, W. Archibald, J. B. Goodenough, Phys. Rev. B, **57**, R2017 (1998).
[8] T. Yoshida, A. Ino, T. Mizokawa, A. Fujimori, Y. Taguchi, T. Katsufuji, Y. Tokura, J. Phys.: Condens. Matter **59**, 258 (2002).
[9] M. D. Banus, T. B. Reed, and A. J. Strauss, Phys. Rev. B **5**, 2775 (1972).
[10] J.-S. Zhou, H. Q. Yin, and J. B. Goodenough, Phys. Rev. B **63**, 184423 (2001).
[11] F. J. Morin, Phys. Rev. Lett. **1**, 34 (1959).
[12] Reaction of vanadium and the silica tube produces $V_3Si$. The presence of small amounts of $V_3Si$, some times even not observable in the x-ray, can be detected as a diamagnetic contribution to the low field susceptibility below 16 K ($V_3Si$ is superconductor below $T_C$=16 K). Surface polishing of the pellets is normally enough to remove any trace of this impurity.
[13] N. F. Mott and H. Jones in *The Theory of the Properties of Metals and Alloys*, Oxford U. P., Oxford, 1936.
[14] M. Morinaga, and J. B. Cohen, Acta Cryst. A**32**, 387 (1976)
[15] J.-S. Zhou, and J. B. Goodenough, Phys. Rev. B **64**, 24421 (2001).
[16] J.-S. Zhou and J. B. Goodenough, Phys. Rev. B **66**, 052401 (2002)
[17] F. Rivadulla, E. Winkler, J.-S. Zhou, J. B. Goodenough, *submitted*.
[18] J. R. Drabble and H. J. Goldsmid, in *Thermal Condution in Semiconductors*, International series of monographs on semiconductors, Pergamon Press, NY, 1961.
[19] C. Kittel, Phys. Rev. **75**, 972 (1949).
[20] G. A. Slack, and R. Newman, Phys. Rev. Lett. **15**, 359 (1958).
[21] D. G. Cahill, and R. O. Pohl, Solid State Comm. **70**, 927 (1989).
[22] L. Kaufman, Trans. Metall. Soc. AIME **224**, 1006 (1962).
[23] J. J. de Yoreo, and R. O. Pohl, Phys. Rev. B **32**, 5780 (1985).